# Conceptual Assessment Tool for Advanced Undergraduate Electrodynamics

Charles Baily[1], Qing X. Ryan[2], Cecilia Astolfi[1] and Steven J. Pollock[3]

[1]School of Physics and Astronomy, University of St. Andrews, St Andrews, Fife KY16 9SS Scotland, UK
[2]Department of Physics and Astronomy, California State Polytechnic University, Pomona, California, 91768 USA
[3]Department of Physics, University of Colorado, Boulder, CO 80309-0390 USA

**Abstract:** As part of ongoing investigations into student learning in advanced undergraduate courses, we have developed a conceptual assessment tool for upper-division electrodynamics (E&M II): the Colorado UppeR-division ElectrodyNamics Test (CURrENT). This is a free response, post-instruction diagnostic with 6 multi-part questions, an optional 3-question pre-instruction test, and accompanying grading rubrics. The instrument's development was guided by faculty-consensus learning goals and research into common student difficulties. It can be used to gauge the effectiveness of transformed pedagogy, and to gain insights into student thinking in the covered topic areas. We present baseline data representing 500 students across 9 institutions, along with validity, reliability and discrimination measures of the instrument and scoring rubric.

**Keywords:** physics education research, upper-division electrodynamics, course transformation, assessment

**PACS:** .40.Fk, 01.40.Di, 01.40.G-, 01.40.gf

## I. INTRODUCTION

Research-validated conceptual assessments play an important role in physics education research (PER). They can be used to document and investigate common student difficulties, and to inform and measure the effectiveness of newly developed curricula [1–3]. Many instruments have been developed for lower-division physics courses [4], but assessing student learning of upper-division content [5–9] presents novel challenges, including: (i) the greater reliance on mathematical formalism in advanced topics, making it more difficult to disentangle conceptual understanding from procedural knowledge; (ii) advanced problem-solving skills cannot be assessed using short-response questions; and (iii) classes tend to be smaller, and are sometimes offered less frequently than large-lecture introductory courses, leading to smaller data sets and less reliable quantitative results.

As part of our multiyear transformation of upper-division physics courses at the University of Colorado Boulder (CU), we developed faculty-consensus learning goals, identified common student difficulties, and designed research-based teaching materials (including clicker questions and in-class tutorials) for the second semester of our junior-level electromagnetism sequence (E&M II) [10–12]. We have also created a validated assessment to quantify and characterize student learning, and to allow for meaningful comparisons of learning outcomes across different academic years, institutions, and teaching methods.

The development of the *Colorado UppeR-division ElectrodyNamics Test* (CURrENT) has been guided by our learning goals and investigations into common student difficulties, and is designed to measure a representative sampling of both procedural knowledge and conceptual understanding within selected core electrodynamics topics (see Griffiths [15], Ch.7-9). Note that a separate instrument, the *Colorado Upper-division Electrostatics* diagnostic (CUE), is available to assess student learning in electrostatics [5, 13].

In this paper, we summarize the development of the CURrENT in section II; present validity, reliability and discrimination studies in sections III, IV & V (respectively); and in section VI we discuss results from an initial data set of 500 students from 9 different institutions.

## II. METHODS AND DEVELOPMENT

### A. Faculty input and learning goals

The Science Education Initiative (SEI) model for course transformation [2] involves three key steps that are used to inform all aspects of the project: (i) establish explicit learning goals in collaboration with experienced faculty; (ii) develop research-informed course materials and teaching strategies to help students achieve these goals; and (iii) use validated assessments to determine what students are (and aren't) learning.

We followed this model by first convening a two-day meeting in summer 2011 of 15 physics faculty members, all with experience in PER and curriculum development, from a total of eight institutions (including CU). Our aim was to brainstorm on student difficulties in advanced undergraduate E&M, and to define our research goals.

We found the coverage of electrostatics to be fairly standard across institutions, but topics from electrodynamics are often treated differently. At CU, electrodynamics is taught in the second half of a two-semester junior-level sequence; classes of 30-60 students meet for three 50-minute periods each week, for a total of 15 weeks. Our usual text is Griffiths (chapters 7-12), though instructors often add topics (e.g., AC circuits) or omit them according to preference. Students at other institutions might instead have just a single semester of advanced undergraduate E&M; use different textbooks; and/or learn about wave optics and relativity in separate courses. To maximize the relevancy of the assessment, we have focused on core material that is likely to be covered in most electromagnetism courses that include time-dependence. It employs fairly standard notation, and in most instances the notation could be readily changed to suit local preferences.

Results from this meeting were supplemented by informal interviews with six instructors who had recently taught electromagnetism at CU. We sought to understand how experienced physicist-teachers had approached this course in the past, ask what they felt were its essential elements, and hear their thoughts on the particular challenges students face when time-dependence is introduced. Our collaborations with non-PER faculty members at CU continued into fall 2011 with 3 informal lunchtime gatherings, to establish explicit learning goals and to vet potential assessment questions.

The SEI method for creating course-scale and topic-specific learning goals is described elsewhere in detail [2, 14]. At the time, course-scale learning goals (regarding students' overall development as physicists) had already been established for electrostatics, and the biggest question was whether our goals for E&M II should differ in any way from those articulated for E&M I. One addition to the list concerns the increasing reliance on mathematics for learning and doing advanced physics. This is particularly true of electrodynamics, which relies heavily on vector calculus, and is typically a student's first encounter with a classical field theory. [See Appendix A for a set of CU faculty consensus learning goals for E&M II.]

Some of these learning goals (such as those related to problem-solving techniques) are already assessed in traditional exams; many others less so, such as articulating correct reasoning, and the CURrENT can be used to measure student attainment in such areas. The questions were designed to assess basic (though not introductory-level) skills and knowledge, with the premise being that a more sophisticated understanding of advanced E&M is unlikely for students who haven't yet mastered these essentials. The learning goals associated with each assessment question are listed in Table 1.

**TABLE 1.** Summary of items on the CURrENT, including point allocations, associated learning goal(s), and Cohen's κ (N=90) [see section IV on reliability]. Question numbers in bold are also on the pretest; starred items are slightly modified in the pretest so as to remove reference to time-dependence.

| Q | Point value | | Description | Goals | κ |
|---|---|---|---|---|---|
| | Pre | Post | | | |
| **1a*** | 2.5 | 5 | Maxwell eqns. (integral) | 2,3 | 0.95 |
| **1b*** | 2.5 | 5 | Vector/surface visualization | 3,4,6,7 | 0.70 |
| **2a*** | 5 | 5 | B field of ∞ solenoid | 1,4,7 | 0.82 |
| 2b | x | 5 | Faraday's law | 1,4,7,8 | 0.91 |
| **3a** | 2.5 | 5 | Translate words into mathematics | 3 | 0.85 |
| **3b** | 2.5 | 5 | Stokes' theorem | 7,10 | 0.78 |
| 4a | x | 5 | Steady-state fields | 6 | 0.91 |
| 4b | x | 5 | Continuity equation | 3,6 | 0.90 |
| 5a | x | 5 | EM field energy | 3,6 | 0.88 |
| 5b | x | 5 | Poynting's theorem | 3,6,9 | 0.77 |
| 6a | x | 2 | Complex exp. notation | 3,6 | 1 |
| 6b | x | 2 | Index of refraction | 1,6 | 1 |
| 6c | x | 2 | Boundary conditions (E) | 2,3,4,6 | 1 |
| 6d | x | 2 | Boundary conditions (B) | 2,3,4,6 | 0.97 |
| 6e | x | 2 | Time-dependence at boundary | 3,6,7 | 0.94 |

### B. Question format and scoring rubric

The post-instruction version of the CURrENT has 6 multi-part questions, with 15 sub-questions, which further break down into a total of 47 scoring elements (the smallest checkpoint where students get a score). The open-response format of most questions allows for the direct assessment of understanding, while simultaneously providing insight into the thinking behind common errors.

The assessment's focus is primarily conceptual, though some mathematical manipulations are required (per our learning goals); for example, Q3 asks students to derive the integral form of a curl equation using Stokes' theorem. More typical of the assessment would be Q4, which asks whether the electric field **E** just outside a current-carrying wire is *zero* or *non-zero*; and likewise regarding the divergence of the steady-state current density **J** inside the wire. [See below and Fig. 1 for further details on Q4.]

To avoid testing computation skills, none of the questions require a numerical response; typical questions ask (e.g.) whether a given quantity is *zero* or *non-zero*, or whether it is *increasing* or *decreasing* with time. In some cases, there are multiple lines of acceptable reasoning for a correct response, which is accounted for in the scoring rubric.

The post-test has a point total of 60. The optional pre-test contains a subset of questions from the post-test, slightly modified so as to be appropriate for pre-instruction, with a point total of 15. As shown in Table 1, each question contributes equally to the total score; similarly, each subpart has equal weighting within a given question. In this paper, all scores are reported as percentages.

Free-response questions require a detailed grading rubric, to explicitly define point allocations for a range of potential responses. Creating an unambiguous rubric can be challenging; usually, a significant amount of training time is necessary in order to achieve reliable scores between different raters [5, 29]. With this in mind, we tried to make scoring the CURrENT as straightforward as possible, to achieve high reliability while minimizing time requirements for both the training and the scoring itself. On average, it takes an experienced rater (i.e., any one of the four authors) roughly 1.5 minutes per student to grade the pretest, and 5 minutes per student for the post-test. [Inter-rater reliability is discussed below in subsection III.B.]

The scoring rubrics include point allocations for each section of a problem, and descriptions of both correct reasoning and common student errors. For example, the problem statement and accompanying scoring rubric for Q4 (a typical question on the CURrENT) are shown in Fig. 1. Each subpart pertains to the same physical scenario, as described in the initial problem statement. Parts (a) and (b) each require unambiguous responses (whether the specified quantity is *zero* or *non-zero*), and for students to articulate the reasoning behind their response. The grading philosophy is essentially all-or-nothing, in that no partial credit is awarded for reasoning that is only partially correct; and there cannot be fully correct reasoning for an incorrect response. Correct responses receive 2 points, and 3 points are given for correctly articulated reasoning. In this case, both subparts each have two distinct lines of reasoning that directly support the correct response.

### C. Administration and data collection

In-class administration is recommended to promote consistent testing conditions. The vast majority of students are able to complete the post-instruction assessment inside a 50-minute class period; the pretest takes approximately 20 minutes or less. The average time for a group of 79 students at CU to complete the post-test was 29 minutes. [Fig. 2.]

Students at CU, and at some of the other institutions, were not informed of the test in advance. It was always administered in class, but not for academic credit, and participation was voluntary. It was always given towards the end of the semester, after the relevant topics had been covered.

---

**4.** A <u>steady</u> current $I_0$ flows in a *long* wire that has a uniform conductivity $\sigma$. The diagram [below] represents the volume current density **J** inside a section of the wire where the diameter is gradually decreasing (the wire itself extends beyond the dotted lines at the ends in the figure).

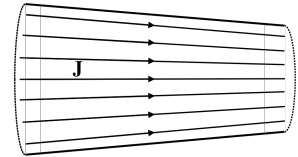

**(a)** Is the electric field **E** just <u>outside</u> the surface of the wire *zero* or *non-zero*? Briefly explain your reasoning.

**Scoring** [5 points total]**:**

[2 points] Indicates the E-field is **non-zero** outside the wire.

[3 points, no partial credit] Correct reasoning; acceptable responses are along the lines of <u>either</u> of the following:

(i) The parallel component of the electric field must be continuous across any boundary. Students do not need to invoke Ohm's law to argue that the electric field is non-zero inside the wire, but they must somehow indicate that the electric field being non-zero inside the wire makes the electric field non-zero outside the wire.

(ii) There are surface charges on the exterior of the wire, which makes the perpendicular component of the electric field non-zero. Students do not need to justify the existence of the surface charges; however, it must be clear they are not claiming that the external E-field is due to a non-zero volume charge density inside the wire. Zero credit given if they claim there would be charges enclosed by some Gaussian surface, but do not explicitly state that these are surface charges.

**(b)** <u>Inside</u> this section of wire, is the divergence of the current density $\nabla \cdot \mathbf{J}$ *zero* or *non-zero*? Briefly explain your reasoning.

**Scoring** [5 points total]:

[2 points] Indicates $\nabla \cdot \mathbf{J} = 0$ inside the wire.

[3 points, no partial credit] Correct reasoning; acceptable responses are along the lines of <u>either</u> of the following:

(i) The current is steady (constant with time), so by the continuity equation (or, conservation of charge) the divergence of the current density must be zero [$\nabla \cdot \mathbf{J} = -\partial \rho / \partial t = 0$]. No need to explicitly invoke the continuity equation. Another acceptable explanation uses Ampere's law for steady currents and the fact that the divergence of a curl is always zero [$\nabla \cdot \mathbf{J} = \nabla \cdot (\nabla \times \mathbf{B} / \mu_0) = 0$].

(ii) The field lines are continuous; they don't start or stop to indicate a "source" or "sink" for the current, making its divergence everywhere zero. E.g., "equal number of lines in and out of any closed surface".

**FIGURE 1.** Problem statement for Q4 of the CURrENT, and related excerpts from the scoring rubric.

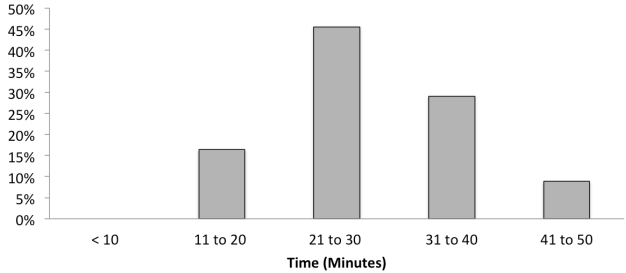

**FIGURE 2.** Histogram of CURrENT completion times for a subset of 79 students at CU Boulder.

The data set reported in this paper includes a total of 500 students from both standard (lecture-based) and transformed courses; at CU-Boulder (five courses) and at eight external institutions (eleven courses). This data set includes only the two most recent iterations of the CURrENT, versions 4 and 5, as there are only slight differences in wording between the two.

This data set is comprised of 39% CU students (N=191), 48% students from 3 other PhD granting institutions (N=237), and 14% students from 5 other Bachelor's degree awarding institutions (N=67). The CU students had a cumulative grade point average (GPA) of 3.3 (out of 4.0) and an average GPA in physics courses of 3.2 prior to taking electrodynamics. We did not ask about the gender, ethnicity or economic background of any student.

## III. VALIDITY

Validity is defined as the extent to which test scores accurately measure the intended concept or construct. We investigate the validity of the CURrENT by asking the following questions: (A) Does the instrument give results that are similar to other measures of the same construct? (B) Do experts agree that the questions are appropriate and fair measures of the associated learning goals? (C) Do students interpret the questions as intended? We show in this section that the CURrENT is, overall, a valid instrument for this population of students in our institutional context.

### A. Criterion Validity

One indicator of validity is the extent to which the test gives results that are similar to other independent measures. We find significant correlations between CURrENT scores and students' final exam performance (Pearson correlation coefficient $r=0.52$, $p<0.001$, $N=276$ including both CU and non-CU students); and also their subject-GPA ($r=0.52$, $p<0.001$, $N=190$, all CU students). These correlations can be characterized as somewhere between ''medium'' (0.3–0.5) and ''strong'' (0.5–1.0) [16], which suggests the constructs measured by the CURrENT are highly related to other aspects of student learning that are valued by faculty.

### B. Expert Validity

The aim of matching individual questions with specific learning goals is to help instructors understand what the assessment is trying to measure, and whether it aligns with their own teaching. We surveyed seven experienced faculty members (two from research institutions and five from 4-year colleges), and one PER researcher, and gave them a copy of the CURrENT and the associated learning goals for each question. They were asked to provide feedback via two guiding questions: (i) Would you expect your students to be able to answer these questions? (ii) Do any of the questions and associated learning goals appear to be inappropriate or mismatched?

All eight stated that they found the questions to be valuable and useful. Four experts proposed small adjustments regarding which learning goals best matched with the questions, and these suggestions were incorporated. Two faculty members had different course structures, where all of junior-level E&M is covered in a single course. We included their opinions to gauge the relevancy of the assessment in a broader sense. Both expressed approval of the content, describing it as "interesting" and "a good cross-section" of the material expected to be covered. One faculty member felt that deriving equations using Stokes' theorem (subpart Q3b, related to learning goal 10) was unimportant to them. "Derivation and proof" is an important learning goal at CU, but in this area it is not surprising that different instructors have different goals.

### C. Student Validation

We conducted approximately 15 student validation interviews during the development of the instrument, to determine whether the questions were being interpreted as intended. Most recently, we interviewed four students using a think-aloud protocol and version 4 of the CURrENT; the interviewer tried not to interject except to remind students to verbalize their thought processes. These interviews were recorded and later analyzed to determine whether student work reflected the intended nature of the question, and whether their written work was consistent with their verbal interpretation of the question. At the conclusion of the interviews, students were asked about items where they had seemed confused, to probe their understanding of the problem statements. A few wording and spacing changes were made as a result; these represent the only significant differences between versions 4 and 5 of the CURrENT.

## IV. RELIABILITY

Whereas validity refers to the extent that scores measure the intended construct, reliability refers to whether the instrument produces similar results under similar conditions. Here, we concentrate on two independent aspects of reliability: (A) Does student performance on any given test item correlate with the remaining items on the test (internal consistency)? (B) How well do different scorers agree with each other on the same student (inter-rater reliability)?

### A. Internal Consistency

Cronbach's alpha (α) is a statistical measure of internal consistency, given by the formula

$$\alpha = \left(\frac{k}{k-1}\right)\left(1 - \sum_{i=1}^{k} \sigma_k^2 / \sigma_t^2\right),$$

where $k$ is the number of test items, $\sigma_t^2$ is the total test variance, and $\sigma_i^2$ is the variance for item $i$. $\alpha$ ranges from 0 to 1, with larger numbers indicating greater internal correlation among test items. Conceptually, if the items on the test are measuring different constructs and are completely independent of each other, one would expect the sum of the variances on individual items to be similar to the variance in the total scores, which leads to a small alpha value.

By treating each subpart (e.g., 1a, 1b, etc.) as individual test items, we obtained α=0.74 (N=500), where α-values between 0.7-0.9 are normally considered adequate [17]. We also computed α more conservatively by treating each question (including all sub parts) as a single test item, and obtained α=0.70. Cronbach's α assumes the unidimensionality of test items [18], and we have no reason to believe the CURrENT measures just a single construct, suggesting our α-values are likely an underestimate of internal consistency.

### B. Inter-rater Reliability

As mentioned above, the items appearing on the CURrENT were written so as to reduce or remove ambiguity regarding what constitutes a correct response, thereby easing the grading process and promoting inter-rater reliability (IRR). We conducted two informal IRR exercises during the development process, wherein: (i) Two of the authors separately graded a subset of five tests and compared scores, looking for ways in which the rubric was lacking. After improvements were made, a second iteration with a different set of five tests was performed, followed by a third iteration. Finally, an instructor from an outside institution was asked to grade his own students independently (N=15), using the existing rubric for version 3 of the assessment and no training from us. We found that the disagreement on the average total score for the class was less than 1%. Rater differences on individual items were typically around 1 or 2%, and were all less than 10% with one exception.

As a more rigorous IRR check, 90 CURrENT exams (all version 4, consisting of two sets: CU (N=47) and a 4-yr college (N=43)) were scored independently by two different raters. Rater 1 (SJP) is a PER faculty member, and rater 2 (QXR) was a PER postdoctoral researcher who had not scored the test previously. The raters initially discussed all the questions and responses for a subset of 11 exams that had been randomly selected from the CU data set, and then scored the rest independently. After the independent scoring, raters discussed only those 8 (out of 47) scoring elements where agreement fell below 90%. The data reported below include the initial training set.

The inter-rater reliability for the two raters working independently was very high; the average total score for the entire data set differed by only 0.2% (0.1% after discussion). We also looked at the absolute value of rater difference to determine the range of variation on individual scores, and this average was 3% (1% after discussion) with a standard deviation of 3% (2% after discussion). Raters agreed on individual total scores to within ±5% for the vast majority of students (86%). Raters differed by more than 10% on only 2% of students (N=2), and the total score for these two students differed by 0% and 2% respectively after discussion.

In this study we also examined inter-rater reliability on individual questions. Rater difference (absolute value) on any individual question averaged 3.5% (1.0% after discussion) and the standard deviation of differences was 9.5% (4.5% after discussion). Similar results were found when comparing scores with a third rater, an undergraduate student researcher (CA), which suggests these findings are robust.

As an additional measure of inter-rater reliability, we computed Cohen's kappa (κ) [19, 20], which is a statistical measure of how often raters give the same score for a question, compared to the proportion expected by random chance; it is defined as

$$\kappa = \frac{\sum f_O - \sum f_E}{N - \sum f_E},$$

where $f_O$ represents the observed frequency of exact agreement, $f_E$ represents the frequency expected by chance, and $N$ is the total number of ratings. We generated a contingency table based on all possible scoring combinations and computed κ (N=90) for all sub-questions. All of our κ values for raters 1 & 2 are "substantial" or better [21; see Table 1], suggesting satisfactory inter-rater agreement.

## V. DISCRIMINATION

We demonstrate in this section that the CURrENT produces: (A) consistent discrimination among students with different levels of understanding; (B) a broad distribution of total scores; and (C) has a reasonable level of item difficulty.

### A. Item-test correlation

We expect that students who score well overall on the test should also tend to score well on individual items. The item-test correlation is typically calculated in terms of point-biserial correlation, but this is only applicable for dichotomous variables. For this open-ended test format, we instead examined the Pearson correlation coefficient for each test item with the rest of the test (with the item itself excluded). The correlation coefficient for each of the six questions ranged from 0.32 and 0.47 [$p < 10^{-12}$ for each]. Minimum acceptable correlation coefficients are generally considered to be around 0.2 [22, 23].

The overall distribution of CURrENT scores is shown in Fig. 3, which also provides a visual indication of the test's discriminatory power, in that there is a good distribution across almost all bins. The normality of the distributions was checked with an Anderson-Darling test for each class (16 total; p>0.05), and for aggregate data from all courses (p<0.05). This means that, with increased sampling, a small variation from the normal distribution could show up as statistically significant. In this case, the average total score is slightly higher than 50% (56.8%), and the tail on the lower end is cut off (no students scored lower than 15%), which could contribute to statistically significant variations from normality in the aggregate score distributions. However, this is not of any practical significance, and is not a cause for concern.

### B. Coefficient of test discrimination

Ferguson's delta (δ), or the "coefficient of test discrimination" [24], measures the discriminatory power of a test by investigating how broadly the total scores of a sample are distributed over the possible range [22].

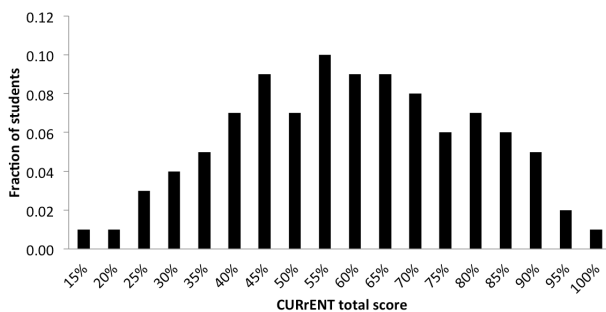

**FIGURE 3.** Histogram of total scores on the CURrENT (N=500).

Calculating Ferguson's δ for a multiple-choice test is straightforward because the number of items and the binning of scores are unambiguous. However, there is not a well-accepted method for calculating δ for open-ended assessments. We used two reasonable alternatives: (i) take the total number of test items (K) as the number of points on the test, and calculate the frequency ($f_i$) of the number of points earned [5]; or (ii) convert the open-ended scoring to multiple choice, simply turning all scoring elements to a corresponding 0 or 1. We obtained δ=0.99 using both methods. The possible range of δ values is [0,1]. Traditionally, δ>0.9 is considered good discrimination; thus the CURrENT can be said to have substantial discriminatory power.

### C. Item Difficulty

The item difficulty index statistic [22] is not applicable because the open-ended scoring is not dichotomous. Instead, we compute the mean for each question to give an idea of the difficulty for each item. As seen in Fig. 4, the questions on the CURrENT are not equally challenging for all students; additional variation is also evident across different student populations. None of the questions yield an extremely high or low percentage, indicating an appropriate level of difficulty for the purposes of discrimination.

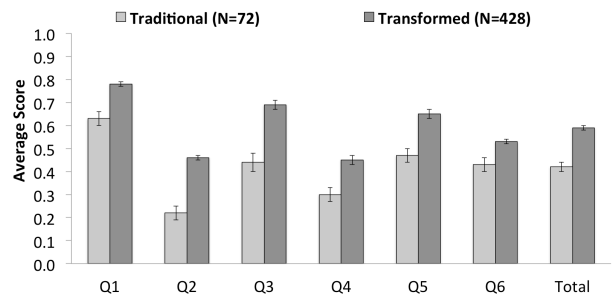

**FIGURE 4.** Comparison of aggregate results on individual questions and total score, for courses that have been characterized as either "traditional" or "transformed" (see text for discussion). Error bars represent the standard error of the mean.

Furthermore, despite the differences between classes/institutions, student cohorts across institutions scored consistently lower on some questions than others (for example, students in 3 courses scored consistently low on Q2, and those in 2 courses scored particularly low on both Q4 & Q5). In other words, some questions are systematically more difficult across multiple populations, suggesting the existence of common students difficulties that should be investigated and addressed.

## VI. RESULTS

### A. Post-test

The average CURrENT post-test score for our data set is 56.8% ±0.9% (N=500). Figure 3 above shows the spread in student performance, ranging from a low of 15% to a high of 100%, following a roughly Gaussian curve. Average total scores for each of the 16 courses included in the data set are shown in Fig. 5.

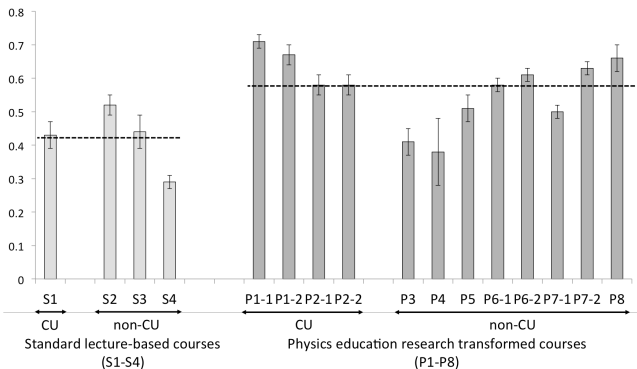

**FIGURE 5.** CURrENT post-test scores across 9 institutions (N=500 students); there were 4 standard lecture courses (S1-S4) and 12 PER transformed courses (P1-P8). Courses are not listed chronologically. Error bars represent the standard error of the mean; the two dashed lines represent averages across the two types of courses, either traditional or transformed (see subsection VI.C for discussion).

### B. Pre-test

The CURrENT pre-test consists of three questions taken from the post-test (as highlighted in Table 1); the questions were modified so as to ensure they would be accessible to students who had completed E&M I (electrostatics). Only some of the courses opted to do both pre- and post-assessment. The average CURrENT pre-test score is 54.3% ±0.3% (N=298). Different populations with differing backgrounds and preparation were seen to have widely different pre-test scores; for example, the average pre-instruction scores for two different non-CU institutions were 63.4% (N=47) and 45.8% (N=17) respectively.

Pre-test scores are found to correlate well with a variety of measures, such as post-test scores on the CUE diagnostic [5] (r=0.59, p<0.001, N=150); and GPA in prior physics courses (r=0.42, p<0.001, N=175). This indicates the pretest score is a meaningful measure of student preparation. Pretest scores are also well correlated with post-test scores (r=0.55, p<0.01, N=298).

The pre-test is primarily useful as an indicator of students' incoming knowledge state, rather than to characterize learning gains over a semester. Not only are the questions not identical, making quantitative comparisons between pre- and post-instruction scores can be problematic. Wallace, et al. [25] have argued that raw scores are typically ordinal (like class rank) but not interval (like calendar years). Ordering students by the number of correct answers allows for a ranking according to the amount of the construct each possesses. [For example, a student who scores a perfect 20 points on an assessment likely knows more about the content than a student who only answered 17 items correctly.] However, we cannot say that a shift from 14 to 17 represents the same increase in knowledge as a shift from 17 to 20. [See [25] for a detailed discussion.]

### C. Broader insights

One difficulty in making meaningful comparisons between pedagogies is that we have far fewer data points in the standard-lecture category. Another challenge lies in characterizing the pedagogies involved. We took a preliminary approach of simply asking instructors whether they would describe their own classroom as being traditional or interactive; if the latter, we asked about the types of interactive engagement they had incorporated into their course. The "transformed" category therefore contains a broad range of interactivity, sometimes mixed in with traditional lecturing. Some instructors (e.g. P1) used many in-class clicker questions (>5 per lecture), as well as tutorials (>=1 per week). Some instructors (e.g. P2) spent relatively less time on clicker questions (1 or 2 per lecture), and tutorials (only a few per semester). Some non-CU instructors used part of the CU materials that were available (e.g. P7 used some CU concept test and homework questions), and some used entirely different approaches (e.g. P6 asked students to view pre-lecture videos before class).

By considering aggregate post-instruction scores from 16 different course offerings, taught by 12 instructors at 9 different institutions, we are able to discern significant differences between traditional versus transformed courses. The four standard lecture-based courses have an average of 42.0%±4.8% (N=72); and courses taught with varying degrees of interactivity had an average of 56.8%±2.9% (N=423). Comparing the average scores of students instead of courses, we have 41.5%±2.0% for the standard lecture-based courses, and 59.3%±0.9% for the transformed courses. [See Fig. 5.]

## VII. DISCUSSION AND CONCLUSIONS

We have developed a free-response conceptual assessment that targets a subset of our faculty-consensus learning goals in electrodynamics. We have achieved a relatively high degree of inter-rater reliability by utilizing question formats that allow for less ambiguity and greater ease of scoring. The validity and reliability of the CURrENT has been evaluated. We find that students generally interpret the questions as intended, and expert opinions were overall positive.

The CURrENT total score is well correlated with other variables, such as final exams and course grades, which are presumably reflections of what instructors value for student learning. The test shows high internal consistency, is able to distinguish between students with different abilities, and reveals measurable differences between different pedagogies. Overall, this instrument shows considerable promise for research and assessment in advanced undergraduate electrodynamics.

The free-response aspect of the CURrENT has proved to be a rich source of insight into student thinking. For example, Q4 has revealed a number of issues with how students physically interpret the divergence of a vector field (see Fig. 1); more specifically, part (b) of this question asks about the divergence of the current density **J** inside a wire carrying a steady current. Because the current is steady, there cannot be any ongoing accumulation of charge anywhere within the wire (no time-dependence), so the continuity equation (an expression of charge conservation) requires the divergence of the current density to be zero everywhere. More than half of students (51.2%, N=284) answered this question incorrectly. [26]

Many students were instead distracted by how the magnitude of **J** increases as the diameter of the wire decreases; or how the density of field lines increases to the right. Students frequently believe this indicates a non-zero divergence in the field, often because they are thinking of the divergence strictly in terms of its differential expression in Cartesian coordinates, and not accounting for contributions from all three components of the field. Other students see the divergence as a global property of a field, rather than local, assuming it is always either everywhere positive, everywhere negative, or zero. These findings have led to classroom interventions that have demonstrably improved student learning in this area (see [26-28] for more details).

We used classical test theory (CTT) [29] for the validation studies. One limitation of CTT is that all test statistics are population dependent, so there is no guarantee that test statistics calculated for one student population (e.g., physics students at a two-year institution) will hold for another population (e.g., physics students at a research focused university).

Item Response Theory (IRT) was specifically designed to address the shortcomings of CTT, in that all item and student parameters are independent of both population and test form. [30, 31] However, a significant limitation in using IRT in the development of upper-division physics assessments is that it requires a large number of statistics (on the order of many hundreds of students, at least). The small class sizes that are typical of upper-division physics courses therefore represent a logistical barrier that is difficult for researchers to overcome. So although IRT could help disentangle student ability from the quality of the test items, the development and validation of upper-division assessments at CU Boulder has always been guided by CTT.

To increase scalability, and to eliminate any remaining scoring ambiguities, a multiple-choice version of the test is a natural next step [32, 33]. The ease of administering and scoring a multiple-choice assessment could help to increase adoption, and consequently the sample size, allowing for more rigorous comparisons between pedagogies, and a more accurate picture of the relative prevalence of student difficulties.

## VIII. ACKNOWLEDGEMENTS

We gratefully acknowledge the many students and staff who made this work possible, in particular: B. Ambrose, P. Beale, A. Becker, T. Drey, M. Dubson, E. Kinney, P. Kohl, F. Kontur, G. Kortemeyer, V. Kumarappan, C. Manogue, D. McIntyre, T. Munsat, R. Pepper, K. Perkins, S. Ragole, E. Redish, D. Rehn, A. M. Rey, C. Rogers, E. Sayre, E. Zimmerman, and the PER@C group. Supported in part by the University of Colorado, the University of St Andrews, and NSF-CCLI grant #1023028.

## APPENDIX A. CU FACULTY-CONSENSUS COURSE-SCALE LEARNING GOALS FOR ELECTRODYNAMICS (E&M II)

These learning goals were created by a group of physics faculty from a number of research areas, including physics education research. Rather than addressing specific content to be covered in a course (as with a syllabus), this list of broader goals represents what we think students should be learning *to do* at this stage of their development as physicists.

1. **Build on earlier material:** Students should deepen their understanding of introductory electromagnetism, junior-level E&M, and necessary math skills (in particular, vector calculus and differential equations).

2. **Maxwell's equations:** Students should see the various topics in the course as part of a coherent theory of electromagnetism; i.e., as a consequence of Maxwell's equations.

3. **Math/physics connection:** Students should be able to translate a description of a junior-level E&M problem into the mathematical equation(s) necessary to solve it; explain the physical meaning of the final solution, including how this is reflected in its mathematical formulation; and be able to achieve physical insight through the mathematics of a problem.

4. **Visualization:** Students should be able to sketch the physical parameters of a problem (e.g., electric or magnetic fields, and charge distributions). They should be able to use a computer program to graph physical parameters, create animations of time-dependent solutions, and compare analytic solutions with computations. Students should recognize when each of the two methods (by hand or computer) is most appropriate.

5. **Organized knowledge:** Students should be able to articulate the important ideas from each chapter, section, and/or lecture, thus indicating how they have organized their content knowledge. They should be able to filter this knowledge to access the information they'll need to solve a particular physics problem, and make connections between different concepts.

6. **Communication.** Students should be able to justify and explain their thinking and/or approach to a problem or analysis of a physical situation, in either written or oral form. Students should be able to understand and summarize a significant portion of an appropriately difficult scientific paper (e.g. an *AJP* article) on a topic from electromagnetism; and have the necessary reference skills to search for and retrieve a journal article.

7. **Problem-solving techniques:** Students should be able to choose and apply the problem-solving technique that is appropriate for a particular situation (e.g., whether to use the integral or differential forms of Maxwell's equations). They should be able to apply these methods to novel contexts (i.e., solving problems that do not map directly to examples in a textbook), indicating how they understand the essential features of the technique, rather than just the rote mechanics of its application.

    …7a. **Approximations:** Students should be able to effectively use approximation techniques, and recognize when they are appropriate (e.g., at points far away or very close to the source). They should be able to decide how many terms of a series expansion must be retained to find a solution of a given order, and be able to complete a Taylor Series to at least two terms.

    …7b. **Symmetries:** Students should be able to recognize symmetries, and be able to take advantage of them when choosing the appropriate method of solution (e.g., correctly applying the Maxwell-Ampere law to calculate the magnetic field of an infinitely long wire).

    …7c. **Integration:** Students should be able to write down the line, surface or volume integral required for solving a specific problem, and correctly follow through with the integration.

    …7d. **Superposition:** Students should recognize that – in a linear system – a general solution can be formed by the superposition of multiple components, and a specific solution found by applying appropriate boundary conditions.

8. **Problem-solving skills:** Students should be able to draw on an organized set of content knowledge (LG#5), and apply problem-solving techniques (LG#7) with that knowledge in order to carry out lengthy analyses of physical situations. They should be able to connect all the pieces of a problem to reach a final solution. They should recognize the value for learning the material of taking wrong turns, be able to recover from their mistakes, and persist in working towards a solution even though they don't necessarily see the path to that solution when they first begin the problem. Students should be able to articulate what it is that needs to be solved for in a given problem, and know when they have found it.

9. **Expecting and checking solutions:** When appropriate for a given problem, students should be able to articulate their expectations for the solution, such as the magnitude or direction of a vector field, the dependence of the solution on coordinate variables, or its behavior at large distances. For all problems, students should be able to justify the reasonableness of a solution (e.g., by checking its symmetry, looking at limiting or special cases, relating to cases with known solutions, dimensional analysis, and/or checking the scale/order of magnitude of the answer).

10. **Derivations/proofs:** Students should recognize the utility and role of formal derivations and proofs in the learning, understanding, and application of physics. They should be able to identify the necessary elements of a formal derivation or proof; and be able to reproduce important ones, including an articulation of their logical progression. They should have some facility in recognizing the range/limitations of a result based on the assumptions made in its derivation.

11. **Intellectual maturity:** Students should accept responsibility for their own learning. They should be aware of what they do and don't understand about physical phenomena and classes of problems, be able to articulate where they are experiencing difficulty, and take action to move beyond that difficulty (e.g., by asking thoughtful, specific questions).